\documentclass[a4paper, amsfonts, amssymb, amsmath, showkeys, nofootinbib, twoside,prd,longbibliography]{revtex4-2}
\usepackage[english]{babel}
\usepackage[utf8]{inputenc}
\usepackage{amsthm}
\usepackage{mathtools}
\usepackage{physics}
\usepackage{xcolor}
\usepackage{graphicx}
\usepackage[left=23mm,right=13mm,top=35mm,columnsep=15pt]{geometry} 
\usepackage{adjustbox}
\usepackage{placeins}
\usepackage[T1]{fontenc}
\usepackage{csquotes}

\usepackage[pdftex, pdftitle={Article}, pdfauthor={Author}]{hyperref} % For hyperlinks in the PDF

\begin{document}
\title{Resonant string behavior in a Gravitational Wave burst}
\author{Vojtěch Liška}
    \email[Correspondence email address: ]{vojta.liska@mail.muni.cz}% Your name
    \affiliation{Department of Theoretical Physics and Astrophysics, Faculty of Science, Masaryk University, Kotlářská 2, 61137 Brno, Czechia}

\author{Rikard von Unge}
    \email[Correspondence email address: ]{unge@physics.muni.cz}% Your name
    \affiliation{Department of Theoretical Physics and Astrophysics, Faculty of Science, Masaryk University, Kotlářská 2, 61137 Brno, Czechia}

\date{\today} % Leave empty to omit a date

\begin{abstract}
We investigate the behavior of classical closed strings in a gravitational wave burst and discover an intriguing resonant behavior where the energy absorbed by the strings is crucially dependent on the amplitude and frequency of the gravitational wave. This behavior can be traced to the well-known behavior of the solutions to the Mathieu equation.
\end{abstract}


\maketitle

\section{Introduction \label{sec:outline}}
    With the detection of gravitational waves \cite{LIGOScientific:2016aoc} we have gained a new channel for information from the universe. Apart from photons and other cosmic ray particles, it is now possible to gather data from gravitational waves. It is therefore of utmost importance to understand how gravitational waves are affected by the matter filling the universe between the place where the gravitational radiation is generated and us. In this paper, we make a modest contribution to this investigation by assuming that the matter consists of classical closed strings and investigating how this affects the gravitational radiation that reaches us.
    
    This topic has been studied before, both at the classical \cite{Amati:1988ww,deVega:1988ts,Costa:1990dy} and quantum \cite{Amati:1988ww,deVega:1988ts,Costa:1990dx,deVega:1990kk,deVega:1991nm,Adamo:2017sze} level. In these papers, the interaction between strings and a sandwich wave \cite{Bondi:1958aj,Penrose:1965rx}, a gravitational wave where the gravitational background differs from flat space only in a small region which is moving with the speed of light in a particular direction. The metric considered was the Aichelburg-Sexl metric, which arises as the shock-wave metric of a black hole moving with the speed of light. In this paper, we instead assume that in the region that differs from Minkowski space, the metric looks like the metric of a gravitational wave of fixed frequency $\omega$ but where the amplitude $A$ is modulated by a Gaussian. We find an intriguing dependence on $A$ and $\omega$ where there are regions in the $A,\omega$ moduli space of strongly resonant behavior and regions with almost no interaction. This could serve as a means of detecting cosmic strings in the universe through gravitational wave astronomy.
    
    The paper is organized as follows: In section \ref{sec:eom} we introduce the equations that we will solve and discuss various gauge choices. The equations are then solved in flat space (section \ref{sec:mink}), in a plane gravitational wave background (section \ref{sec:plane}), for a delta function, sandwich wave (section \ref{sec:sw}) and finally for the gravitational wave burst (section \ref{sec:Num}). A detailed discussion about the method used to compute the energy of the string after being hit by the wave is given in an appendix. We end up with conclusions containing a discussion of future work.
    
    All plots and numerical computations were performed using the Julia language \cite{bezanson2017julia}.

\section{The String equations \label{sec:eom}}
We want to study the behavior of strings when meeting a gravitational wave burst, which we will describe using the Brinkman metric \cite{Brinkman:1923,Brinkman:1925}
\begin{align}
    ds^2 = -2du dv + H(u,x,y) du^2 + dx^2 +  dy^2 \; .
\end{align}
In order for the metric to be a vacuum solution, the function $H$ has to satisfy the Laplace equation
\begin{align}
    (\partial_x^2 + \partial_y^2) H = 0 \; ,
\end{align}
whereas the $u$ dependence of $H$ is completely arbitrary. The standard gravitational waves are described by choosing $H = (x^2-y^2)f(u) + 2xy g(u)$. The two terms correspond to the two polarizations of the gravitational wave and $f(u)$ and $g(u)$ are usually chosen to be harmonic functions.

Starting from the Nambu-Goto action \cite{Nambu:1986ze,Goto:1971ce} for a string in a nontrivial background $G_{MN}(X)$
\begin{align}
    S  = \int d\tau d\sigma \sqrt{-\det\left(\partial_a X^M\partial_b X^N G_{MN}\right)} \;,
\end{align}
we find the equations of motion
\begin{align}
\partial_a\left(\sqrt{-\gamma}\gamma^{ab}G_{MN}\partial_b X^N\right) -
\frac12 \sqrt{-\gamma}\gamma^{ab}\partial_M G_{NK} \partial_a X^N
\partial_b X^K = 0 \;,
\end{align}
where $\gamma_{ab} = \partial_a X^M\partial_b X^N G_{MN}$ is the pull-back metric on the world sheet.

The Nambu-Goto action is reparametrization invariant and to solve the equations we need to fix a gauge, i.e., to choose the coordinates $\tau$ and $\sigma$ on the world sheet. This can be done so that $\gamma_{\tau\sigma} = 0$ and $\gamma_{\tau\tau} +\gamma_{\sigma\sigma} = 0$, so-called orthogonal or conformal gauge.

With this gauge choice in the gravitational wave background, the equations that we need to solve are
\begin{align}
\label{ueq}
    (\partial_\tau^2-\partial_\sigma^2)u &= 0 \; , \\
    (\partial_\tau^2-\partial_\sigma^2)x &= \frac{\partial_x H}{2}\left[(\partial_\tau u)^2-(\partial_\sigma u)^2\right] \; , \\
    (\partial_\tau^2-\partial_\sigma^2)y &= \frac{\partial_y H}{2}\left[(\partial_\tau u)^2-(\partial_\sigma u)^2\right] \;, \\
    (\partial_\tau^2-\partial_\sigma^2) v &=
    \frac{\partial_u H}{2}\left[(\partial_\tau u)^2-(\partial_\sigma u)^2\right]
    +\partial_x H(\partial_\tau {u}\partial_\tau x-\partial_\sigma u \partial_\sigma x)
    +\partial_y H(\partial_\tau{u}\partial_\tau y-\partial_\sigma u \partial_\sigma y)
    \label{veq}
    \; ,
\end{align}
supplemented by the gauge choice
\begin{align}\label{eq:gort}
    \partial_\tau u \partial_\sigma v + \partial_\tau v \partial_\sigma u &= H\partial_\tau u \partial_\sigma u + \partial_\tau x \partial_\sigma x +\partial_\tau y \partial_\sigma y \;, \\
    2\partial_\tau u\partial_\tau v + 2\partial_\sigma u \partial_\sigma v &= H((\partial_\tau u)^2 + (\partial_\sigma u)^2)+
    (\partial_\tau x)^2 +(\partial_\sigma x)^2 +
    (\partial_\tau y)^2 +(\partial_\sigma y)^2\label{eq:glen} \; .
\end{align}
The gauge is implemented by first choosing the $\tau$ coordinate proportional to $u$
\begin{align}\label{eq:uch}
    u &= \lambda \tau \;,
\end{align}
where $\lambda$ is an arbitrary constant. With this choice, the equations that ensure that $\partial_\tau$ and $\partial_\sigma$ are orthogonal (\ref{eq:gort}) and of equal length (\ref{eq:glen}) simplify
\begin{align}
    \label{vseq}
    \lambda \partial_\sigma v &= 
    \partial_\tau x \partial_\sigma x +\partial_\tau y \partial_\sigma y \;, \\
    \label{vteq}
    2\lambda\partial_\tau v &= \lambda^2 H+
    (\partial_\tau x)^2 +(\partial_\sigma x)^2 +
    (\partial_\tau y)^2 +(\partial_\sigma y)^2 \;,
\end{align}
and the equations for the transverse coordinates also simplify to become
\begin{align}
    \label{xeq}
    (\partial_\tau^2-\partial_\sigma^2)x &= \frac{\lambda^2}{2}{\partial_x H} \;,\\
    \label{yeq}
    (\partial_\tau^2-\partial_\sigma^2)y &= \frac{\lambda^2}{2}{\partial_y H} \;.
\end{align}
It is straightforward to check that (\ref{eq:uch}), (\ref{vseq}) and (\ref{vteq}) imply (\ref{veq}).

\section{Flat space\label{sec:mink}}
As a warmup exercise we will solve the equations for the case $H=0$ which is equivalent to Minkowski space. To begin with, we solve (\ref{xeq}) and (\ref{yeq}). The simplest way to find a set of solutions is to separate variables so that if $x(\tau,\sigma) = T_x(\tau) S_x(\sigma)$ we get
\begin{align}
    \frac{\partial_\tau^2 T_x}{T_x} =\frac{\partial_\sigma^2 S_x}{S_x} = -K \;.
\end{align}
The $2\pi$ periodicity in $\sigma$ then tells us that we have to choose $K = n^2$ with $n$ being an integer. The equation for $y$ is solved similarly. The simplest nontrivial solution in this family is
\begin{align}\label{eq:flat}
    x &= R \cos(\tau)\cos(\sigma) \;, \\
    y &= R \cos(\tau)\sin(\sigma) \;,
\end{align}
with $R$ constant.
Plugging this into (\ref{vseq}) and (\ref{vteq}) we get that
$\partial_\sigma v = 0$ and $\partial_\tau v = \frac{R^2}{2\lambda}$. In the more familiar $t$ and $z$ coordinates, this corresponds to
\begin{align}
    t &= \frac{1}{\sqrt{2}}(u+v) = (\lambda + \frac{R^2}{2\lambda})\frac{\tau}{\sqrt{2}} \;, \\
    z &= \frac{1}{\sqrt{2}} (u - v) = (\lambda - \frac{R^2}{2\lambda})\frac{\tau}{\sqrt{2}} \;.
\end{align}
We see that in general, the $z$-coordinate changes as a function of $t$ corresponding to a motion of the center of mass of the string. However, if we choose $\lambda = \frac{R}{\sqrt{2}}$ the $z$ coordinate is constant and there is no center of mass motion in the $z$ direction but a different choice of $\lambda$ would include also this possibility.

\section{A string in a plane gravitational wave \label{sec:plane}}
We now turn to the topic of the motion of the string in a plane gravitational wave background described by
\begin{align}
    H = (x^2-y^2)A \cos(\omega u) \;,
\end{align}
where $A$ is the amplitude and $\omega$ is the frequency of the wave. The equation for the transversal coordinates now becomes
\begin{align}
    (\partial_\tau^2-\partial_\sigma^2) x &=
    \lambda^2 A \cos(\omega u) x \;, \\
    (\partial_\tau^2-\partial_\sigma^2) y &=
    -\lambda^2 A \cos(\omega u) y \;.
\end{align}
Again we try separation of variables to find a solution. This leads to the equations
\begin{align}
    \frac{\partial_\tau^2 T_x}{T_x} - \lambda^2 A \cos(\lambda\omega \tau) = \frac{\partial_\sigma^2 S_x}{S_x} = -n^2 \;, \\
    \frac{\partial_\tau^2 T_y}{T_y} + \lambda^2 A \cos(\lambda\omega \tau) = \frac{\partial_\sigma^2 S_y}{S_y} = -n^2 \;,
\end{align}
from which we can extract the equations for $T_x$ and $T_y$
\begin{align}
    \partial_\tau^2 T_x + (n^2 - \lambda^2 A \cos(\lambda\omega \tau)) T_x &= 0 \;,
    \\
    \partial_\tau^2 T_y +(n^2 + \lambda^2 A \cos(\lambda\omega \tau) ) T_y &= 0 \;,
\end{align}
and we see that $T_x$ and $T_y$ now satisfies the Mathieu equation
\begin{align}
    \frac{d^2 w(t)}{d t^2} +(a-2q\cos(2t))w(t) = 0 \;,
\end{align}
with
\begin{align}
    a &= \frac{4n^2}{\lambda^2\omega^2} \;, \\
    q &= \pm \frac{2A}{\omega^2} \;,
\end{align}
and with the Mathieu functions as solutions.

\begin{figure}[htb]
\includegraphics[width = 8.6cm]{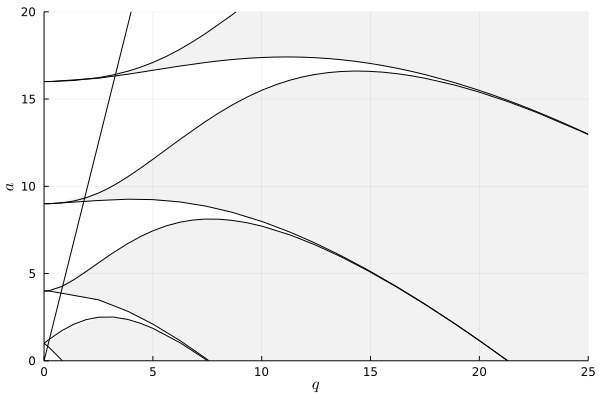}
\caption{The moduli space of the Mathieu equation. The shaded regions correspond to unstable solutions. The parameters $a$ and $q$ vary on a straight line for a gravitational wave with arbitrary frequency $\omega$ but with fixed amplitude $A$.}
\label{fig:MatStab}
\end{figure}

The behavior of the Mathieu functions depending on the parameters $a$ and $q$ is quite intricate. As can be seen in figure \ref{fig:MatStab}, there are regions in the $(q,a)$ plane where the solutions, although not periodical, behave nicely and stay finite for all times. There are other regions where instead the solutions oscillate uncontrollably as time goes to infinity. On the borders between these regions, the solutions are periodic.

We see that the type of solution we get will depend crucially on the amplitude and frequency of the gravitational wave. If our choice of parameters is such that the solution is in an unstable region, our string will exhibit resonant behavior where it will fluctuate with an ever larger amplitude as well as start moving in the $z$-direction.

To see this, we have to investigate the behavior of the $v$ coordinate which tells us what the string is doing in the $z$ direction. For simplicity, we choose $n=1$. Starting with (\ref{vseq}) the $\sigma$ derivative of $v$ is
\begin{align}
    \lambda\partial_\sigma v = \frac{1}{4}\partial_\tau(T_y^2-T_x^2)\sin(2\sigma) \;,
\end{align}
with the simple solution
\begin{align}
    \label{vtemp}
    v = \frac{1}{8\lambda}\partial_\tau(T_x^2-T_y^2)\cos(2\sigma) + k(\tau) \;,
\end{align}
where $k(\tau)$ is an arbitrary function. In the case that $T_x^2-T_y^2$ is not a constant $v$ depends on $\sigma$, and since $z=\frac{1}{\sqrt{2}}(u-v)$ this means that the string will oscillate also in the $z$ direction even if it does not from the beginning. To fix the unknown function $k(\tau)$ we insert (\ref{vtemp}) into (\ref{vteq}) to get an expression for the $\tau$ derivative of $k(\tau)$
\begin{align}
    \frac{dk}{d\tau} =
    \frac{1}{8\lambda}\left[
    \frac{d^2}{d\tau^2}( T_x^2 +  T_y^2)
    +4(T_x^2 + T_y^2)\right] \;.
\end{align}
 It is a nontrivial consistency check that the $\sigma$ dependence of this equation drops out. The function $k(\tau)$ encodes the center of mass motion in the $z$-direction. For instance, if $k\neq \frac{R^2}{2\lambda}\tau$, $u\neq v$ and the center of mass of the string is moving in the $z$-direction.

If we would like to find the full motion of the string in this background, we must choose an initial configuration. We could, for instance, choose the configuration that we studied in flat space where the string starts out as a circle with radius $R$ and no motion in the $z$ direction. That means we have to choose the initial values (for example at $\tau = 0$)
\begin{align}
    T_x &= R \;, \\
    \dot T_x &= 0 \;, \\
    T_y &= R \;, \\
    \dot T_y &= 0 \;, \\
    \lambda &= \frac{R}{\sqrt{2}} \;.
\end{align}
Solving the equations for the fluctuation in the $x,y$ plane, we find
\begin{align}
    T_x(\tau) &= R C_{a,q}\left(\frac{R\omega\tau}{\sqrt{2}}\right) \;, \\
    T_y(\tau) &= R C_{a,-q}\left(\frac{R\omega\tau}{\sqrt{2}}\right) \;,
\end{align}
where $C_{a,q}(x)$ is the even solution to the Mathieu differential equation normalized so that $C(0)=1$, which gives the full solution
\begin{align}
    x & = R C_{a,q}\left(\frac{R\omega\tau}{\sqrt{2}}\right) \cos\sigma \;,
    \\
    y &= R C_{a,-q}\left(\frac{R\omega\tau}{\sqrt{2}}\right) \sin\sigma \;,
    \\
    v &= \frac{R}{4\sqrt{2}}\partial_\tau(C_{a,q}^2-C_{a,-q}^2)\cos(2\sigma) + k(\tau) \;,
    \\
    u &= \frac{R}{\sqrt{2}}\tau \;.
\end{align}
One can easily convince oneself that $C_{a,q}^2-C_{a,-q}^2$ is not constant in general and therefore the string vibrates also in the $z$-direction. Also, for $a$ and $q$ in the unstable region, the amplitude of the vibrations in the $x,y$ plane increases without bound.

\section{A string in a sandwich wave\label{sec:sw}}
We would like to study the more realistic situation when the motion of the string starts and ends in Minkowski space, but there is a short gravitational wave burst during which the string may gain (or lose) energy from the gravitational wave. An exactly solvable example is when the gravitational wave is a delta function
\begin{align}
    H = (x^2-y^2)A\delta(u) \;,
\end{align}
which is known as the Aichelburg-Sexl metric and the behavior of strings in this background has been studied extensively both in the classical \cite{Amati:1988ww,deVega:1988ts,Costa:1990dy} and quantum \cite{Amati:1988ww,deVega:1988ts,Costa:1990dx,deVega:1990kk,deVega:1991nm} cases. We do not claim to derive any new results in this section. Rather, we use some of the techniques developed here to study a more realistic case in chapter \ref{sec:Num}.

In the case at hand, the equations that need to be solved are
\begin{align}
    (\partial_\tau^2-\partial_\sigma^2)x = \lambda^2 A \delta(u) x \;, \\
    (\partial_\tau^2-\partial_\sigma^2)y = -\lambda^2 A \delta(u) y \;,
\end{align}
and a separation of variables leads to
\begin{align}
    \partial_\tau^2 T_x +(n^2-\lambda^2 A \delta(\lambda\tau))T_x = 0 \;, \\
    \partial_\tau^2 T_y +(n^2+\lambda^2 A \delta(\lambda\tau))T_y = 0 \;.
\end{align}
Integrating these equations from $\tau = -\epsilon$ to $\tau = \epsilon$ where $\epsilon$ is infinitesimal, we find
\begin{align}
    \partial_\tau T_x(\epsilon) = \partial_\tau T_x(-\epsilon) +
    \lambda A T_x (0) \;, \\
    \partial_\tau T_y(\epsilon) = \partial_\tau T_y(-\epsilon) 
    - \lambda A T_y (0) \;.
\end{align}
In particular, if we assume that the string motion before meeting the wave is given by the flat space solution (\ref{eq:flat}), for which $n=1$. Then, using the continuity of the string, the motion after the wave hits is given by
\begin{align}
    T_x = R\left(\cos(\tau) +\lambda A \sin(\tau)\right) \;, \\
    T_y = R\left(\cos(\tau) - \lambda A \sin(\tau)\right) \;.
\end{align}
Integrating the equations for $v$, (\ref{vseq}) and (\ref{vteq}), gives
\begin{align}
    v = \frac{R^2}{2\lambda}\left((1+\lambda^2A^2)\tau + \lambda A \cos(2\tau) \cos(2\sigma)\right) \;.
\end{align}
If the string was at rest before meeting the wave, we know that $\lambda = \frac{R}{\sqrt{2}}$. The full solution after meeting the wave is then
\begin{align}\label{delx}
    x &= R\left(\cos(\tau) +\frac{A}{\sqrt{2}} \sin(\tau)\right)\cos(\sigma) \;, \\
    \label{dely}
    y &= R\left(\cos(\tau) - \frac{A}{\sqrt{2}} \sin(\tau)\right)\sin(\sigma)\;, \\
    \label{delv}
    v &= \frac{R}{\sqrt{2}}\left(\left(1+\frac{R^2A^2}{2}\right)\tau + \frac{R A}{\sqrt{2}} \cos(2\tau) \cos(2\sigma)\right)\;, \\
    \label{delu}
    u &= \frac{R}{\sqrt{2}}\tau \;.
\end{align}
We recognize that the center of mass of the sting has started moving in the negative $z$-direction with the speed $\frac{A^2R^2}{4+R^2 A^2}$ but it has also started vibrating in the $z$-direction. To disentangle the motion, it is useful to make a Lorentz transformation to a frame in which the center of mass of the string is at rest. In that system 
\begin{align}
    v &= \frac{R}{\sqrt{2}}\left(\sqrt{1+\frac{R^2A^2}{2}}\tau + \frac{R A}{\sqrt{2}\sqrt{1+\frac{R^2A^2}{2}}} \cos(2\tau) \cos(2\sigma)\right) \;, \\
    u &= \frac{R}{\sqrt{2}}\sqrt{1+\frac{R^2A^2}{2}}\tau \;.
\end{align}

We find the internal energy of the string by using a method outlined in the appendix. It involves going to static gauge which can be found through a conformal transformation. The outcome in this case is
\begin{align}
    E_{int} = \sqrt{ 1 +\frac{R^2 A^2}{2} } E_i \;,
\end{align}
where $E_i$ is the energy of the initial string configuration.
From the point of view of an observer at rest with respect to the string before meeting the gravitational wave, it looks like the string has increased its mass (internal energy) and has started moving in the $z$-direction. The total increase in energy combines these two effects and is therefore
\begin{align}
    E = \left( 1 +\frac{R^2 A^2}{4} \right) E_i \;.
\end{align}

\section{A String in a gravitational wave burst\label{sec:Num}}
To study a more realistic example, we combine the previous cases and assume a gravitational wave profile
\begin{align}\label{eq:burst}
    H(x,y,,u) &= (x^2-y^2) f(u) \;, \\
    f(u) &= A \cos(\omega u) e^{-\frac{u^2}{\rho^2}} \;,
\end{align}
where the constant $\rho$ was chosen such that the burst lasts for approximately ten periods. This is the simplest approximation of what a real gravitational wave that we observe on Earth could look like. However, since our calculations are numerical, there is nothing that prevents us from using an even more realistic waveform if we should like.

\begin{figure}[htb]
\includegraphics[width = 8.6cm]{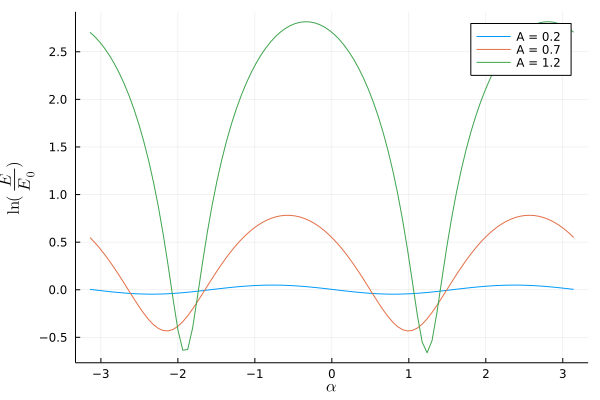}
\caption{The energy increase as a function of the relative phase $\alpha$ of the incoming gravitational wave burst and the string.}
\label{fig:relphase}
\end{figure}

Even though in this case we cannot solve the equations for the transverse directions analytically, we know that they have to take the form
\begin{align}
    x &= \sum_{n=-\infty}^{\infty}c_n(\tau)e^{in\sigma} \;, \\
    y &= \sum_{n=-\infty}^{\infty}d_n(\tau)e^{in\sigma} \;,
\end{align}
which, with the function $H$ specified by (\ref{eq:burst}) leads to the equations
\begin{align}
    \ddot{c}_n + (n^2-\lambda^2 f(u))c_n &= 0 \;, \\
    \ddot{d}_n + (n^2+\lambda^2 f(u))c_n &= 0 \;,
\end{align}
for the unknown coefficients, $c_n$ and $d_n$. At the same time, we make the ansatz
\begin{align}
    v = \sum_{n=-\infty}^{\infty}v_n(\tau)e^{in\sigma} \;,
\end{align}
which gives the equations
\begin{align}
    \lambda n v_n &= \sum_{k} k (c_k \dot{c}_{n-k} + d_k \dot{d}_{n-k}) \;,
\end{align}
for $n\neq 0$. From this equation, we find that the higher $v_n$ modes in general get excited. For instance, If $c_{\pm 1}$ and $d_{\pm 1}$ are nonzero we should expect that $v_{\pm 2}\neq 0$. In the previous section, where we studied the delta function sandwich wave, the $c$ and $d$ contributions canceled before the arrival of the wave burst but not after the burst had passed as can be clearly seen in (\ref{delv}). The center of mass motion in the $z$-direction is encoded in $v_0$ which, in contrast to the other $v_n$ ($n\neq0$) modes, has to be integrated from
\begin{align}
    \dot{v}_0 &= \frac{1}{2\lambda}\sum_k\left( \dot{c}_k\dot{c}_{-k} +\dot{d}_k\dot{d}_{-k}
    +k^2c_k c_{-k} +k^2 d_{k}d_{-k} + \lambda^2A \cos\lambda\omega\tau
    \left( c_k c_{-k} -d_k d_{-k}\right)\right) \;.
\end{align}
Following a procedure completely analogous to what we did for the delta function wave burst, we first find a Lorentz frame in which the center of mass of the string is at rest followed by a conformal transformation to static gauge. It is then straightforward to find the energy of the string. We perform these manipulations numerically and find that the amount of energy that the string absorbs depends strongly on the amplitude $A$ and the frequency $\omega$ of the wave burst. What is surprising is that the energy absorbed also depends on the relative phase of the vibrations of the string and the vibrations of the gravitational wave. As can be seen in figure \ref{fig:relphase}, for particular choices of relative phase, the string might even lose energy.
Because of this, it seems reasonable to average of the relative phase, if the gravitational wave burst would travel through a gas of strings, the relative phases would be random.

\begin{figure}[hbt]
\includegraphics[width = 8.6cm]{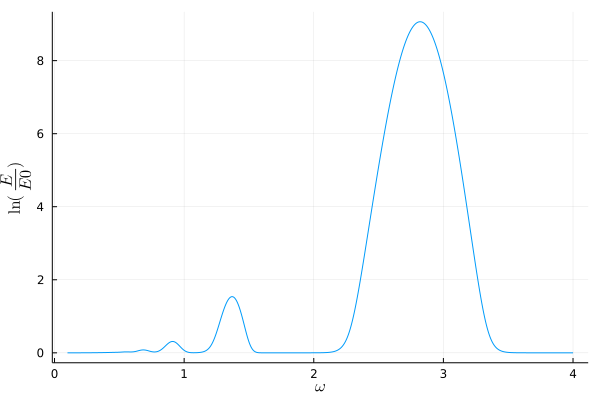}
\caption{The energy increase for fixed $A$ as a function of $\omega$.}
\label{fig:spek}
\end{figure}

Doing this, we find that the energy absorption is still strongly dependent on the amplitude $A$ and the frequency $\omega$ of the gravitational wave burst. This is illustrated in figure \ref{fig:spek} where for a constant $A = 1.5$ we look at the increase in energy as a function of $\omega$. Notice that the energy scale is logarithmic so that at resonance the energy increases by several orders of magnitude in the resonance region.
The peaks in the spectrum can be matched with the unstable regions of the Mathieu equation in the $a,q$-plane. For instance, when the initial string is at rest so that $\lambda = \frac{R}{\sqrt{2}}$, and for a fixed value of $A$. Changing $\omega$ changes $a$ and $q$ along a straight line in the $a,q$-plane, as can be seen in figure \ref{fig:MatStab}, according to the relation
\begin{align} \label{eq:Rdep}
    a &= k q \;, \\
    k &= \frac{4n^2}{R^2A} \;, \\
    q &= \frac{2A}{\omega^2} \;.
\end{align}
Here $a=q=0$ corresponds to $\omega = \infty$ but lowering $\omega$ we move a way from the center until we get to the unstable region. For $R=1$ and $A=1.5$ (and $n=1$) we numerically find that this happens when $\omega \approx 3.33$ and continuing to lower $\omega$ we get out of the unstable region at $\omega \approx 2.29$. These values agree precisely with the spectrum in figure \ref{fig:spek} showing that the resonance has its origin in the behavior of the Mathieu equation. Changing $A$ we change the slope of the line $a= k q$ so that for a smaller $A$ the line moves towards the $y$-axis giving a smaller interval with resonant behavior, whereas for a bigger $A$, the line moves towards the $x$-axis and the resonance bands become broader. It is interesting to note that the center of the interval does not change significantly during this since $k\cdot q$ is independent of $A$.

\begin{figure}[hbt]
\includegraphics[width = 8.6cm]{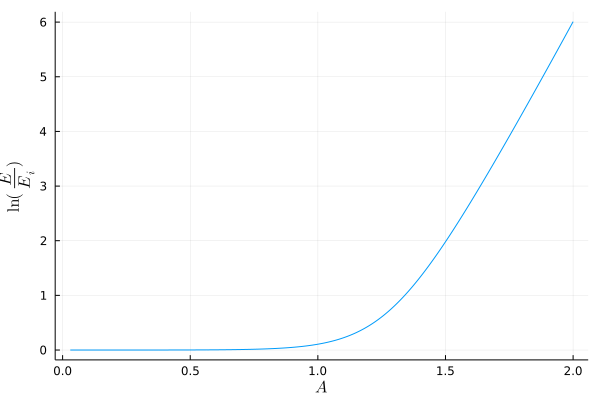}
\caption{The energy increase for fixed $\omega$ as a function of $A$}
\label{fig:EA}
\end{figure}

On the other hand, choosing a fixed frequency $\omega$ at a value where there is resonance, and plotting the increase in energy as a function of the amplitude as in figure \ref{fig:EA}, we see that the absorption increases exponentially after an initial period of slow increase.

\begin{figure}[hbt]
\includegraphics[width = 8.6cm]{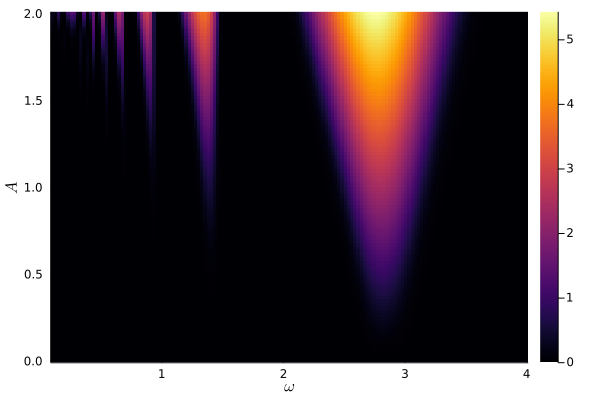}
\caption{The energy increase as a function of the frequency $\omega$ and amplitude $A$ of the gravitational wave burst.}
\label{fig:AoE}
\end{figure}

More generally, we give the dependence on both $A$ and $\omega$ in figure \ref{fig:AoE}
Notice that the energy scale is logarithmic, so the increase in energy can be several orders of magnitude.

Finally, it is necessary to discuss the choice of units. One might expect the string tension to appear and it does indeed in the calculation of the energy of the string. However, since we are only interested in the relative increase in energy, the string tension actually drops out. For the other parameters, $A$ has dimension length$^{-2}$ whereas $\omega$ has dimension length$^{-1}$. Looking at (\ref{eq:Rdep}) we see that our results only depend on the dimensionless combinations $R^2A$ and $R\omega$. Using a string in an initial state with a different value of $R$ would just rescale the values of $A$ and $\omega$ accordingly.

\section{Conclusions\label{sec:conc}}
We have shown that the behavior of classical closed strings in a gravitational wave burst is strongly dependent on the amplitude and frequency of the gravitational wave. The behavior can be traced to the known behavior of the Mathieu equation. Our study is restricted to the simplest possible case where the string fluctuates in the lowest mode only and the fluctuations are transverse to the direction of the gravitational wave. In a more realistic situation, the orientation of the string would be arbitrary and in general, more modes would be excited, possibly according to a Boltzmann distribution. It is interesting to observe that the presence of a mode in the initial state gives rise, through (\ref{vteq}), to a mode of twice the frequency in the final state. This could presumably give an even stronger resonance behavior than the case treated in this paper. Another straightforward modification would be to modify the wave form of the gravitational wave burst. In principle, one could even use the waveform measured by LIGO defined numerically. We leave all of this for future publications.

We have shown that strings are able to absorb significant amounts of energy from the gravitational wave. The nontrivial frequency dependence indicates that this could modify the spectrum of gravitational waves as observed here on Earth. A fundamental string in flat space would be expected to have an $R$ of the order of the Planck length $R \propto l_p = 10^{-35}m$ and the resonance region of $\omega$ would then be given in Planck units. To move the resonance to a region of frequencies measurable by LIGO or Virgo ($10^1 - 10^4$ Hz) \cite{Abbott:2016xvh,VIRGO:2010okc}, one would have to have a string of size $R \propto 10^{-4} m$. This would be possible for cosmic strings with the appropriate tension. For fundamental strings, one would need some other mechanism to make them large. For instance, as Susskind has pointed out, a highly excited string, on the verge of becoming a black hole, spreads out to large size \cite{Susskind:1993aa,Susskind:1993ws} thus lowering the resonance frequency, but this would require a more thorough analysis.

\section*{Acknowledgements} \label{sec:acknowledgements}
We would like to thank Ulf Lindström and Linus Wulff for useful discussions. The work of Rikard von Unge is supported by the Czech science foundation GA\v{C}R through the grant “Integrable Deformations” (GA20-04800S).

\appendix*
\section{Calculating the energy}\label{sec:appendix}
Here we explain how to find the $z$-velocity and the energy of the string after meeting the gravitational wave. In all the cases studied in this article, the form of the solution of the coordinates $u$ and $v$ can be written as
\begin{align}
    u &= \lambda\tau \;, \\
    v &= \lambda M^2 \tau + M(g(\tau+\sigma) + g(\tau-\sigma) \;,
\end{align}
where $M$ is a constant and $g$ is an arbitrary function. Through the equations of motion (\ref{vseq}) and (\ref{vteq}) we can also compute that
\begin{align}
    \partial_\tau x \partial_\sigma x + \partial_\tau y \partial_\sigma y &=
    \lambda \partial_\sigma v = M(g^\prime (\tau + \sigma) - g^\prime(\tau - \sigma)) \;, \\
    \partial_\tau x \partial_\tau x + \partial_\sigma x \partial_\sigma x +
    \partial_\tau y \partial_\tau y + \partial_\sigma y \partial_\sigma y &=
    2\lambda \partial_\tau v = \lambda M^2 + M(g^\prime (\tau + \sigma) + g^\prime(\tau - \sigma)) \;.
\end{align}

After a Lorentz transformation in the $u,v$ plane with velocity $v_z = \frac{1-M^2}{1+M^2}$ we find
\begin{align}
    u &= \lambda M \tau \;, \\
    v &= \lambda M \tau + g(\tau + \sigma) + g(\tau - \sigma) \;,
\end{align}
which implies that
\begin{align}
    t &= \frac{1}{\sqrt{2}}\left(2\lambda M\tau + g(\tau+\sigma) + g(\tau - \sigma)\right) \;, \\
    z &= -\frac{1}{\sqrt{2}}\left(g(\tau+\sigma) + g(\tau - \sigma)\right) \;.
\end{align}

We would like to compute the internal energy of the string using the method described in Chapter 7 of \cite{Zwiebach:2004tj} where in static gauge, in which we will call the world-sheet coordinates $\mathfrak{t}$ and $\mathfrak{s}$, we can find the energy as
\begin{align}\label{eq:E}
    E = \int d\mathfrak{s} \sqrt{\frac{(\partial_\mathfrak{s}\bar{X})^2}{1-\bar{v}_{\perp}^2}} \;,
\end{align}
where $\bar{X} = (x,y,z)$ and
\begin{align}
    \bar{v}_\perp = \partial_\mathfrak{t} \bar{X} - \frac{\partial_\mathfrak{t}\bar{X}\cdot\partial_\mathfrak{s}\bar{X}}{(\partial_\mathfrak{s}\bar{X})^2} \partial_\mathfrak{s}\bar{X} \;,
\end{align}
is the local velocity of the string perpendicular to the string itself.

To go to static gauge, we make a conformal transformation
\begin{align}
    \mathfrak{t}\pm \mathfrak{s} 
    = \sqrt{2}\left(\lambda M (\tau \pm \sigma) + g(\tau \pm \sigma)\right) \;,
\end{align}
and then, inverting the relation
\begin{align}
    \left(\begin{array}{c} \partial_\tau \bar{X} \\ \partial_\sigma \bar{X}
    \end{array}\right) =
    \left(\begin{array}{cc}
    \frac{\partial \mathfrak{t}}{\partial \tau}   & \frac{\partial \mathfrak{s}}{\partial \tau}\\
    \frac{\partial \mathfrak{t}}{\partial \sigma}   & \frac{\partial \mathfrak{s}}{\partial \sigma}
    \end{array}\right)
    \left(\begin{array}{c} \partial_\mathfrak{t}\bar{X} \\ \partial_\mathfrak{s}\bar{X}
    \end{array}\right) \;,
\end{align}
we can express the integral (\ref{eq:E}) in terms of known variables. After a non-trivial calculation, one can show that the internal energy of the string can be expressed as
\begin{align}
    E = \int d\mathfrak{s} =  \sqrt{2}2\pi \lambda M
\end{align}
Combining this with the boost, the total energy gained by the string is
\begin{align}
    E_{tot} = \frac{1+M^2}{2} E_i
\end{align}


%\bibliography{GWBString}
\end{document}